\documentclass[a4paper,11pt]{article}
\usepackage{pos}
\usepackage{bm}
\usepackage{tikz}
\usepackage{pgfplots}
\usetikzlibrary{intersections}
\usepgfplotslibrary{fillbetween,colorbrewer}
\pgfplotsset{
	compat=newest,
	width=7cm,
	ylabel style={rotate=-90},
	xmajorticks=false,
	ymajorticks=false,
	axis x line=center,
	axis y line=left,
	legend cell align=left,
	reverse legend,
	colormap/Dark2,
	cycle multiindex* list={
		Dark2\nextlist
		linestyles\nextlist
		very thick
	},
}

\pgfmathsetmacro{\chiquarkonium}{0.096}
\pgfmathsetmacro{\chihybrid}{0.012}
\pgfmathsetmacro{\stringtension}{0.948}
\pgfmathsetmacro{\aone}{-0.341}
\pgfmathsetmacro{\atwo}{0.368}
\pgfmathsetmacro{\bone}{0.353}
\pgfmathsetmacro{\btwo}{-37.589}
\pgfmathsetmacro{\quarkoniumoffset}{-0.228}
\pgfmathsetmacro{\hybridpioffset}{0.565}
\pgfmathsetmacro{\hybridsigmaoffset}{0.573}
\pgfmathsetmacro{\mixstrength}{0.095}
\pgfmathsetmacro{\mixoffset}{-1.0}
\pgfmathsetmacro{\hbar}{197.3}
\pgfmathsetmacro{\keight}{0.037}
\pgfmathsetmacro{\sommer}{0.5}
\pgfmathsetmacro{\mpi}{0.6}

\title{The Pattern of Exotic Hidden-Heavy Hadrons Revealed}

\author*{Roberto Bruschini}

\affiliation{Department of Physics, The Ohio State University,\\
  191 W Woodruff Ave, Columbus, Ohio 43210, USA}

\emailAdd{bruschini.1@osu.edu}

\abstract{%
For more than twenty years, theory has failed to explain the pattern of the exotic heavy hadrons. We illustrate a simple solution to this longstanding puzzle using the Born-Oppenheimer approximation for QCD. Exotic hidden-heavy hadrons are bound states and resonances in potentials that are repulsive at short range and cross a heavy-hadron--pair threshold before approaching it. This explains the proximity of the exotic hidden-heavy hadrons to heavy-hadron--pair thresholds, identifies the thresholds that support bound states or resonances, and prevents an explosion in the number of predicted states. We also discuss the fine tunings of QCD that are responsible for the remarkable properties of some of the exotic hidden-heavy mesons.%
}

\FullConference{The 21st International Conference on Hadron Spectroscopy and Structure (HADRON2025)\\
 27 - 31 March, 2025\\
Osaka University, Japan\\}

\begin{document}
\maketitle

\section{Introduction}

Exotic hidden-heavy hadrons have puzzled theoretical physicists for more than 20 years, and hundreds of papers have been written about them \cite{Ali17,Esp17,Leb17,Guo18,Kar18,Ols18,Bra20,Che23}.
Most studies investigate the nature of the exotic hidden-heavy hadrons using constituent models.
Quark and diquark models assume that they are composites of quarks and/or diquarks interacting through the strong nuclear force.
Molecular models assume that they are bound states of pairs of hadrons interacting through light-hadron exchange.
Constituent models have many parameters whose relation to the fundamental theory QCD is unknown.
Other studies calculate the properties of the exotic hidden-heavy hadrons using lattice QCD and the L\"uscher formalism.
However, these calculations are made extremely difficult by the presence of decay channels with multiple hadrons in the final state.
Another approach that is firmly based on QCD is the Born-Oppenheimer (B\nobreakdash-O) approximation.
In the B\nobreakdash-O approximation for QCD, hidden-heavy hadrons are bound states and resonances in potentials that correspond to the energy levels of QCD with static triplet and antitriplet color sources separated by a variable distance.
This approach has been formulated as an effective field theory for all hadrons containing two heavy quarks or antiquarks \cite{Ber15,On17,Bra18a,Sot20a,Ber24}.

In spite of these overwhelming efforts, theory has so far failed to reveal the pattern of the exotic hidden-heavy hadrons.
The nature of the known states has not been firmly understood and, while the debate rages on, more states are being discovered by collider experiments.
This embarrassing situation might lead to the somewhat nihilistic conclusion that there is no simple solution to the problem.
But is looking for a simple explanation of the exotic hidden-heavy hadrons really a lost cause?

Here we illustrate a simple solution to the problem of the exotic hidden-heavy hadrons from Reference~\cite{Braa24b}.
In Section~\ref{sec:pattern}, we reveal the pattern of the exotic hidden-heavy hadrons using the B\nobreakdash-O approximation for QCD.
In Section~\ref{sec:tetraquarks}, we apply this pattern to hidden-heavy tetraquarks and we calculate some of their properties using simple models for their B\nobreakdash-O potentials.
Finally, we summarize these results in Section~\ref{sec:summary}.

\section{The Pattern}
\label{sec:pattern}

\subsection{The Born-Oppenheimer Approximation for QCD}

An exotic hidden-heavy hadron contains a heavy quark ($Q$), a heavy antiquark ($\bar{Q}$), light quarks and antiquarks, and gluons.
The mass of the heavy quark and antiquark $m_Q$ is typically much larger than the energy scale associated with the gluons and light quarks $\Lambda_\text{QCD}$.
The B\nobreakdash-O approximation for QCD amounts to treating the $Q$ and $\bar{Q}$ as ``slow'' degrees of freedom whose motion is adiabatic with respect to the evolution of the ``fast'' degrees of freedom, that is, the light quarks and gluons.
This allows the calculation of the spectrum of exotic hidden-heavy hadrons to be reduced to a two-step procedure:
\begin{enumerate}
\item Calculate B\nobreakdash-O potentials $V(r)$ using lattice QCD with static triplet and antitriplet color sources separated by a variable distance $r$.
\item Solve a Schr\"odinger equation for the $Q\bar{Q}$ pair in those B\nobreakdash-O potentials.
\end{enumerate}

The B\nobreakdash-O potentials are labeled by quantum numbers for the symmetries of QCD with static triplet and antitriplet color sources separated by a vector $\bm{r}$.
The continuous symmetries are rotations around $\bm{\hat{r}}$, which are generated by $\bm{J}\cdot\bm{\hat{r}}$ where $\bm{J}$ is the angular momentum vector.
The discrete symmetries are $CP$ and reflections $R$ through a plane containing $\bm{\hat{r}}$.
The traditional notation for the B\nobreakdash-O quantum numbers is $\Lambda_\eta^\epsilon$ where $\Lambda$ is $\Sigma,\Pi,\Delta,\dots$ for $\lvert \bm{J}\cdot\bm{\hat{r}} \rvert = 0,1,2,\dots$, the subscript $\eta$ is $g,u$ for $CP=+,-$, and the superscript $\epsilon$ is $+,-$ for $R=+,-$.
If $\Lambda$ is not $\Sigma$, the superscript $\epsilon$ is omitted since states with $\epsilon=+,-$ are just equal-energy superpositions with opposite $\bm{J}\cdot\bm{\hat{r}}$.

The masses of the hidden-heavy hadrons and their decay widths into heavy-hadron pairs can be calculated by solving a B\nobreakdash-O Schr\"odinger equation for bound states and resonances.
We write the B\nobreakdash-O Schr\"odinger equation in the diabatic representation,
\begin{equation}
-\frac{1}{m_Q}\nabla^2 \Psi(\bm{r}) + \mathbf{V}(\bm{r}) \Psi(\bm{r}) = E \Psi(\bm{r}),
\label{eq:schr}
\end{equation}
where $\Psi(\bm{r})$ is a multichannel wavefunction and $\mathbf{V}(\bm{r})$ is the \emph{diabatic potential matrix}; see Reference~\cite{Bru23a} and references therein.
Each element of the $\bm{r}$-dependent diabatic potential matrix matrix $V(\bm{r})$ is a linear combination of B\nobreakdash-O potentials $V_{\Lambda_\eta^\epsilon}(r)$ that depend only on $r=\lvert \bm{r}\rvert$ with coefficients given by Wigner $D$-functions of the polar angles $\theta$ and $\phi$.

\subsection{Potentials in Pure SU(3) Gauge Theory}

The earliest calculation of a B\nobreakdash-O potential for QCD was that of the ground-state potential in pure SU(3) gauge theory \cite{Sta84}.
Excited potentials in pure SU(3) gauge theory were calculated much later \cite{Jug99,Jug03}.
At large $r$, the potentials approach a linear confining potential associated with a gluonic string stretching between the triplet and antitriplet color sources, labeled by the excitation level $n$ of the string.
At small $r$, these potentials approach a color-Coulomb potential offset by a constant, where they form degenerate multiplets labeled by quantum numbers $J^{PC}$.
The development of potential nonrelativistic QCD revealed that these small-$r$ multiplets of potentials are associated with discrete states of QCD with a single color source \cite{Bra00}.

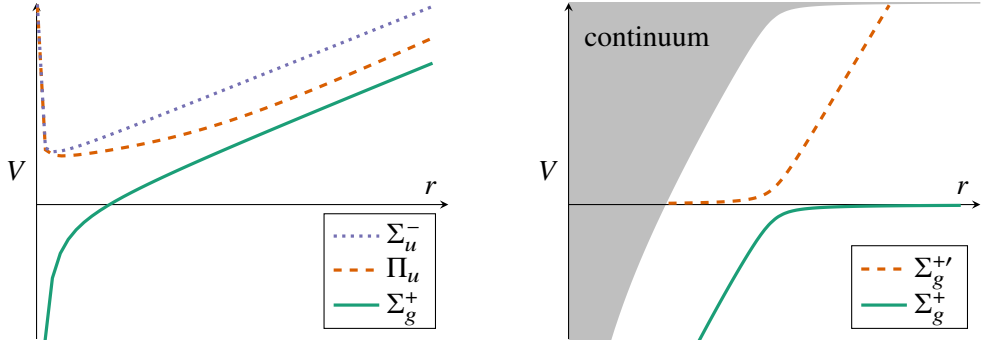
\begin{figure}
	\centering
	\begin{tikzpicture}
		\begin{axis}[
			xlabel={$r$},
			ylabel={$V$},
			legend style={legend pos=south east},
			xmin = 0,
			xmax=2.6,
			ymin=-2,
			ymax=3
		]
			\addplot+[
				samples=50,
				domain=0.05:2.5
			]{
				-\chiquarkonium / x + \stringtension * x  + \quarkoniumoffset
			};
			\addplot+[
				samples=50,
				domain=0.005:2.5
			]{
				\chihybrid / x  * (1 + \bone * x + \btwo * x ^ 2) / (1 + \aone * x + \atwo * x ^ 2) + \stringtension * x + \hybridpioffset
			};
			\addplot+[
				samples=50,
				domain=0.005:2.5
			]{
				\chihybrid / x  + \stringtension * x + \hybridsigmaoffset
			};
			\legend{$\Sigma_g^+$, $\Pi_u$, $\Sigma_u^-$}
		\end{axis}
		\begin{axis}[
			at={(7cm, 0)},
			xlabel={$r$},
			ylabel={$V$},
			legend style={legend pos=south east},
			fill between/on layer=axis background,
			xmin=.1,
			xmax=2.2,
			ymin=-.4,
			ymax=.6
		]
			\addplot+[
				samples=50,
				smooth,
				domain=.1:2.1,
			]{
				(-\chiquarkonium / x + \stringtension * x + \mixoffset) / 2 - sqrt((-\chiquarkonium / x + \stringtension * x + \mixoffset) ^ 2 / 4 + \mixstrength ^ 2 / 4)
			};
			\addplot+[
				samples=50,
				smooth,
				domain=.61:1.735,
				]{(-\chiquarkonium / x + \stringtension * x + \mixoffset) / 2 + sqrt((-\chiquarkonium / x + \stringtension * x + \mixoffset) ^ 2 / 4 + \mixstrength ^ 2 / 4)
			};
			\addplot[
				draw=none,
				samples=50,
				smooth,
				domain=.1:2.2,
				name path=A
			]{
				(-\chiquarkonium / x + \stringtension * x + \mixoffset) / 2 - sqrt((-\chiquarkonium / x + \stringtension * x + \mixoffset) ^ 2 / 4 + \mixstrength ^ 2 / 4) + \mpi
			};
			\addplot[
				draw=none,
				domain=.1:2.2,
				name path=B]{.8};
	    	\addplot[gray, opacity=.5] fill between[of=A and B];
			\node at (0.5,0.5) {continuum};
			\legend{$\Sigma_g^+$,$\Sigma_g^{+\prime}$,,,}
		\end{axis}
	\end{tikzpicture}
	\caption{\label{fig:bopots}Qualitative behavior of the lowest B-O potentials $\Sigma_g^+$, $\Pi_u$, and $\Sigma_u^-$ in pure SU(3) gauge theory (left) and the two lowest $\Sigma_g^+$ B\nobreakdash-O potentials in QCD with light quarks (right).}
\end{figure}

The qualitative behavior of the lowest B\nobreakdash-O potentials $\Sigma_g^+$, $\Pi_u$, and $\Sigma_u^-$ in pure SU(3) gauge theory is displayed in the left panel of Figure~\ref{fig:bopots}.
The lowest potential is $\Sigma_g^+$, which approaches the attractive potential for the vacuum $0^{++}$ at small $r$ and the linear potential for a ground-state string $n=0$ at large $r$.
Bound states in the $\Sigma_g^+$ potential are associated with conventional quarkonium mesons.
The next two lowest potentials are $\Pi_u$ and $\Sigma_u^-$, which approach the repulsive potential for the ground-state gluelump $1^{+-}$ at small $r$ and the linear potentials for a singly and triply excited string $n=1,3$ at large $r$.
Bound states in the $\Sigma_u^-$ and $\Pi_u$ potentials are associated with quarkonium hybrid mesons.

\subsection{Heavy-Hadron Pair Potentials}

The ground-state potential $\Sigma_g^+$ has a very different behavior in QCD with two light quarks.
As $r$ increases, the $\Sigma_g^+$ potential crosses over from a linear potential to a constant equal to twice the energy of the ground-state $\bm{3}$-meson, which consists of a light quark bound to a static color-triplet source.
The $\Sigma_g^+$ potential has a narrow avoided crossing with an excited $\Sigma_g^{+\prime}$ potential that, conversely, crosses over from a constant to a linear potential with increasing $r$ \cite{Bal05}.
The qualitative behavior of the two lowest $\Sigma_g^+$ B\nobreakdash-O potentials in QCD with light quarks is displayed in the right panel of Figure~\ref{fig:bopots}.

These results showed that the B\nobreakdash-O potentials for hidden-heavy hadrons include heavy-hadron--pair potentials and they must approach those potentials at large $r$.
It is however difficult to establish the small-$r$ behavior of the heavy-hadron--pair potentials using lattice QCD.
The difficulty arises from the fact the excited $\Sigma_g^{+\prime}$ potential at small $r$ crosses into the continuum region starting at the energy of the ground-state $\Sigma_g^+$ potential plus the pion mass; see the right panel of Figure~\ref{fig:bopots}.

\subsection{Adjoint-Hadron Potentials}

Although the heavy-hadron--pair potentials at small $r$ are difficult to calculate using lattice QCD, they can be constrained in the limit $r\to0$ using first principles.
At $r\to0$, the triplet and antitriplet color sources behave as a single color source that is a superposition of singlet and octet.
Therefore, discrete B\nobreakdash-O potentials must approach a color-Coulomb potential at small $r$ offset by a discrete energy of QCD with a single color source.
The color-Coulomb potential is attractive or repulsive if the color source is singlet or octet, respectively.

Since a localized singlet color source is the same as no source at all, its spectrum includes only one discrete state: the vacuum with $J^{PC}=0^{++}$.
This corresponds to the lowest $\Sigma_g^+$ potential at small $r$, which is an attractive color-Coulomb potential.
On the other hand, QCD with a localized octet color source has an infinite number of discrete states that are called \emph{adjoint hadrons}.
Each adjoint hadron is associated with a multiplet of \emph{adjoint-hadron potentials} at small $r$.

In pure SU(3) gauge theory, the lowest-energy adjoint-hadron is the gluelump with $J^{PC}=1^{-+}$.
It corresponds to the repulsive $\Pi_u$ and $\Sigma_u^-$ potentials at small $r$ in the left panel of Figure~\ref{fig:bopots}.
In QCD with light quarks, the lowest-energy adjoint hadrons also include two adjoint mesons with $J^{PC}=1^{--}$ and $0^{-+}$ \cite{Fos99}.
The $1^{--}$ adjoint meson corresponds to repulsive $\Sigma_g^+$ and $\Pi_g$ potentials at small $r$.
The $0^{-+}$ adjoint meson corresponds to a repulsive $\Sigma_u^-$ potential at small $r$.

\subsection{Potentials for Exotic Hidden-Heavy Hadrons}

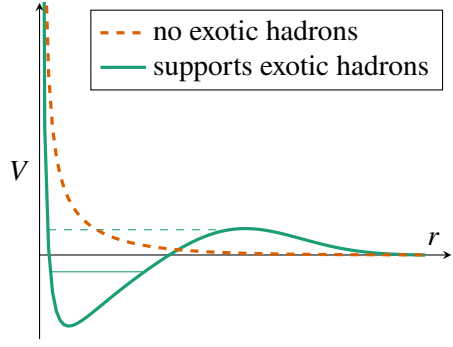
\begin{figure}
	\centering
	\begin{tikzpicture}
		\begin{axis}[
			xlabel={$r$},
			ylabel={$V$},
			xmin = 0,
			xmax = 3.2,
			ymin=-50,
			ymax=150
		]
				\addplot+[
					samples=100,
					domain=0.01:3,
					name path=potential
				]{
					(\keight * \hbar / x - 100) * exp(-x / \sommer) + 20* exp(-(x - 1.5)^2 / \sommer)
				};
				\path[name path=bound state]
					(0,-10) -- (3, -10);
				\path[name path=resonance]
					(0,15) -- (3, 15);
				\draw[index of colormap=0, name intersections={of=potential and bound state}]
					(intersection-1) -- (intersection-2);
				\draw[dashed, index of colormap=0, name intersections={of=potential and resonance}]
					(intersection-1) -- (intersection-2);
				\addplot+[
					samples=100,
					domain=0.01:3
				]{
					(\keight * \hbar / x + 20) * exp(-x / \sommer)
				};
				\legend{supports exotic hadrons,no exotic hadrons}
		\end{axis}
	\end{tikzpicture}
	\caption{\label{fig:exopot}An adjoint-hadron potential can approach a heavy-hadron--pair potential either by decreasing monotonically (dashed curve) or by crossing the threshold before approaching it (solid curve). In the latter case, it may support bound states and resonances (thin horizontal lines).}
\end{figure}

The spectrum of QCD with static triplet and antitriplet color sources must be a smooth function of their distance $r$.
Therefore, the B\nobreakdash-O potentials for exotic hidden-heavy hadrons must necessarily connect smoothly adjoint-hadron potentials at small $r$ to heavy-hadron--pair potentials at large $r$.
The simplest possibility is that the potential approaches the threshold from above, in which case it would not support bound states.
Another simple possibility is that it crosses below the threshold before approaching it, in which case it may support bound states and resonances.
These two possibilities are displayed in Figure~\ref{fig:exopot}.

We identify the exotic hidden-heavy hadrons with bound states and resonances in potentials that are repulsive at small $r$, cross a heavy-hadron--pair threshold at intermediate $r$, and approach that threshold at large $r$.
This identification provides a qualitative explanation of the energies and the number of exotic hidden-heavy hadrons.
Since their potentials are repulsive at small $r$ and constant at large $r$, the energies of the exotic hidden-heavy hadrons must necessarily lie close to a heavy-hadron--pair threshold.
Such potentials can only admit a finite number of bound states and resonances, and only the potentials associated with the lowest-energy adjoint hadrons are likely to cross below the threshold for a heavy-hadron pair.
This requires that the spectrum of exotic hidden-heavy hadrons is limited, preventing an explosion in the number of predicted states.

\section{Application to Hidden-Heavy Tetraquarks}
\label{sec:tetraquarks}

\subsection{Model Potentials}

We now apply the proposed pattern to hidden-heavy tetraquarks using simple models for the B\nobreakdash-O potentials $\Sigma_g^+$ and $\Pi_g$ associated with a $1^{--}$ adjoint meson and $\Sigma_u^-$ associated with a $0^{-+}$ adjoint meson.
Our model for the $\Lambda_\eta^\epsilon$ potential associated with a $J^{PC}$ adjoint meson with isospin $I$ is
\begin{equation}
V_{\Lambda_\eta^\epsilon}^{(I)}(r) =
\begin{cases}
\kappa_8 / r +  E_{J^{PC}}^{(I)} + A_{\Lambda_\eta^\epsilon}^{(I)}\, r^2  &
\text{if $r < R_{\Lambda_\eta^\epsilon}^{(I)}$,} \\
B_{\Lambda_\eta^\epsilon}^{(I)}\,e^{-r/d}  &
\text{if $r > R_{\Lambda_\eta^\epsilon}^{(I)}$.}
\end{cases}
\label{eq:param}
\end{equation}
The value of $B_{\Lambda_\eta^\epsilon}^{(I)}$ is determined by requiring continuity at the matching radius $R_{\Lambda_\eta^\epsilon}^{(I)}$.
The value of $R_{\Lambda_\eta^\epsilon}^{(I)}$ is determined by requiring smoothness at the matching radius.
The strength of the color-Coulomb potential is the same for all potentials, $\kappa_8=0.037$.
For the relaxation length $d$ in the exponential we choose the value $0.5$~fm.
We take the values of $A_{\Lambda_\eta^\epsilon}^{(I)}$ from the parametrizations B\nobreakdash-O potentials in pure $\mathrm{SU(3)}$ gauge theory in Reference~\cite{Ala24}: $0.88\text{ fm}^{-3}$, $9.44 \text{ fm}^{-3}$, and $5.44\text{ fm}^{-3}$ for $\Sigma_g^+$, $\Pi_g$, and $\Sigma_u^-$, respectively.
We treat the adjoint-meson energies $E_{J^{PC}}^{(I)}$ as adjustable parameters.

The B\nobreakdash-O potentials at leading order in $1/m_Q$ have exact heavy-quark spin symmetry (HQSS).
Thus, each bound state corresponds to a HQSS multiplet of hidden-heavy tetraquarks.
At first order in $1/m_Q$, there are spin-dependent corrections to the B\nobreakdash-O potentials that split the energies of the tetraquarks in a HQSS multiplet.
At short distance, the spin-dependent correction include the spin-splitting of adjoint-mesons.
At large distance, the spin-dependent correction include the spin-splitting of well-separated $\bm{3}$-mesons and $\bm{\bar{3}}$-mesons.
The correction at large distances is a constant spin-splitting term that depends on the spins $\bm{S}_Q$ and $\bm{S}_{\bar{Q}}$ of the heavy quark and antiquark and the angular momenta $\bm{j}_{\bm{3}}$ and $\bm{j}_{\bm{\bar{3}}}$ of the light QCD fields inside the $\bm{3}$-meson and $\bm{\bar{3}}$-meson.
For the lowest heavy-meson--pair potentials, the spin splitting term is
\begin{equation}
V_\text{SS} = \Delta \bm{j}_{\bm{3}} \cdot \bm{S}_Q + \Delta \bm{j}_{\bm{\bar{3}}} \cdot \bm{S}_{\bar{Q}},
\end{equation}
where $\Delta$ is a parameter of order $1/m_Q$ that can be fixed to the spin splitting between the ground-state heavy mesons with $J^P=1^-$ and $0^-$.

\subsection{Isospin-0 Tetraquarks}

We identify $\chi_{c1}(3872)$ \cite{Cho03} with the ground state in B\nobreakdash-O potentials associated with a low-energy, isospin-0 adjoint meson.
Only the $1^{--}$ adjoint meson is relevant to the quantum numbers $J^{PC}=1^{++}$ of $\chi_{c1}(3872)$.
We take the charm quark mass to be $m_c = 1.48$~GeV.
We calculate the critical adjoint-meson energy for which the ground state is exactly at threshold as $-157$~MeV.

The $\chi_{c1}(3872)$ is the $1^{++}$ member of a HQSS quartet whose other members are $0^{++}$, $2^{++}$, and $1^{+-}$.
We identify the $1^{--}$ adjoint-meson energy with its critical energy, so the energy of the multiplet is the spin-weighted average $D^{(\ast)}\bar{D}^{(\ast)}$ threshold, which we take as 0.
We calculate the spin splittings of the hidden-charm tetraquarks using first-order perturbation theory in the spin-splitting potential $V_\text{SS}$ from Equation~\eqref{eq:param}.
The energies of the four states are 
\begin{equation}
E_{0^{++}} = -\Delta, \qquad
E_{1^{++}} = -\tfrac{1}{2} \Delta, \qquad
E_{2^{++}} = +\tfrac{1}{2} \Delta, \qquad
E_{1^{+-}} = 0.
\end{equation}
We take the value of $\Delta$ from the experimental splitting between $D^\ast$ and $D$ mesons, $\Delta=141$~MeV.
The $1^{++}$ and $2^{++}$ states are bound states near the $D^\ast\bar{D}$ and $D^\ast\bar{D}^\ast$ thresholds, respectively.
The $0^{++}$ and $1^{+-}$ states are resonances halfway between the $D\bar{D}$ and $D^\ast\bar{D}$ thresholds and halfway between the $D^\ast\bar{D}$ and $D^\ast\bar{D}^\ast$ thresholds, respectively.

\subsection{Isospin-1 Tetraquarks}

We identify $T_{b\bar{b}1}(10610)$ and $T_{b\bar{b}1}(10650)$ \cite{BELLE12} with ground states in B\nobreakdash-O potentials associated with two low-energy, isospin-1 adjoint mesons.
Both the $1^{--}$ and $0^{-+}$ adjoint mesons are relevant to the quantum numbers $J^{PC}=1^{+-}$ of $T_{b\bar{b}1}(10610)$ and $T_{b\bar{b}1}(10650)$.
We take the bottom quark mass to be $m_b = 4.89$~GeV.
We calculate the critical adjoint-meson energies for which the ground state is exactly at threshold as $-95$~MeV for the $1^{--}$ adjoint meson and $-107$~MeV for the $0^{-+}$ adjoint meson.

The $T_{b\bar{b}1}(10610)$ and $T_{b\bar{b}1}(10650)$ are orthogonal superpositions of two $1^{+-}$ states belonging to different HQSS multiplets.
One is a heavy-quark spin singlet ($S_{Q\bar{Q}}=0$) state in a HQSS quartet whose other members are $0^{++}$, $1^{++}$, and $2^{++}$.
The other is a heavy-quark spin triplet ($S_{Q\bar{Q}}=1$) state in a HQSS doublet whose other member is $0^{++}$.
We identify the $1^{--}$ and $0^{-+}$ adjoint-meson energies with their critical energies, so the two multiplets are degenerate and their energy is the spin-weighted average $B^{(\ast)}\bar{B}^{(\ast)}$ threshold, which we take as 0.
We calculate the spin splittings of the hidden-bottom tetraquarks using first-order perturbation theory in the spin-splitting potential $V_\text{SS}$ from Equation~\eqref{eq:param}.
The energies of the six states are 
\begin{equation}
E_{0^{++}} = -\tfrac{3}{2}\Delta, \quad
E_{0^{++\prime}} = +\tfrac{1}{2}\Delta, \quad
E_{1^{++}} = -\tfrac{1}{2} \Delta, \quad
E_{2^{++}} = +\tfrac{1}{2} \Delta, \quad
E_{1^{+-}} = -\tfrac{1}{2} \Delta, \quad
E_{1^{+-\prime}} = +\tfrac{1}{2} \Delta. \quad
\end{equation}
We take the value of $\Delta$ from the experimental splitting between $B^\ast$ and $B$ mesons, $\Delta=45$~MeV.
All six states are near-threshold bound states.
The $0^{++}$ state is near the $B\bar{B}$ threshold.
The $1^{++}$ and $1^{+-}$ states are near the $B^\ast\bar{B}$ threshold.
The $0^{++\prime}$, $1^{+-\prime}$, and $2^{++}$ states are near the $B^\ast\bar{B}^\ast$ threshold.
We also calculate the probability for the various states to have $S_{Q\bar{Q}}=0$ or $S_{Q\bar{Q}}=1$.
The $1^{++}$ and $2^{++}$ states are pure $S_{Q\bar{Q}}=1$.
The two $0^{++}$ states are orthogonal superpositions of $S_{Q\bar{Q}}=0$ and $S_{Q\bar{Q}}=1$ with probabilities $\frac{1}{4}$ and $\frac{3}{4}$.
The two $1^{+-}$ states, which we identify with $T_{b\bar{b}1}(10610)$ and $T_{b\bar{b}1}(10650)$, are orthogonal superpositions of $S_{Q\bar{Q}}=0$ and $S_{Q\bar{Q}}=1$ with equal probabilities.

A remarkable feature of $T_{b\bar{b}1}(10650)$ is that its decays into $B^\ast\bar{B}$ are not observed although they are kinematically favored over the observed decay mode $B^\ast\bar{B}^\ast$.
This suppression is explained by $T_{b\bar{b}1}(10650)$ being an equal-probability superposition of $S_{Q\bar{Q}}=0$ and $S_{Q\bar{Q}}=1$ that is orthogonal to the heavy-quark spin state of $B^\ast\bar{B}$.
Thus, the B\nobreakdash-O approximation reproduces the result obtained by Voloshin using HQSS and light-quark spin symmetry \cite{Vol16}.

\section{Summary}
\label{sec:summary}

We have identified the exotic hidden-heavy hadrons as bound states and resonances in repulsive potentials at short distances that cross below a heavy-hadron--pair threshold before approaching it at large distances.
This simple proposal explains the proximity of their masses to heavy-hadron--pair thresholds, prevents an explosion in the number of predicted states, and identifies the fine tunings of QCD that are responsible for their amazing properties.
The pattern of the exotic hidden-heavy hadron is largely determined by the spectrum of adjoint hadrons in QCD.
This realization provides a powerful incentive to calculate the masses of adjoint hadrons using lattice QCD.

Until calculations of their B\nobreakdash-O potentials in lattice QCD are available, the best one can do is use model potentials for the exotic hidden-heavy hadrons.
Here we have calculated some properties of hidden-heavy tetraquarks using simple model potentials with spin splittings included using first-order perturbation theory.
More accurate predictions of the masses and decay widths can be obtained by including the effects of spin splittings nonperturbatively.
The verification of such predictions will reveal that the pattern of the exotic hidden-heavy hadrons has finally been understood.

\acknowledgments

This work was supported in part by the U.S. Department of Energy under grant DE-SC0011726.
I acknowledge support from Osaka University for travel to the conference.
I would like to thank E. Braaten for useful discussions.
This work contributes to the goals of the US DOE ExoHad Topical Collaboration, Contract DE-SC0023598.

\bibliographystyle{JHEP}
\bibliography{bibliography}

\end{document}